\documentclass[twocolumn,twocolappendix]{aastex63}

\usepackage{todonotes}
\usepackage{cleveref}
\usepackage{booktabs}
\usepackage{longtable}
\usepackage{array}

\makeatletter
\usepackage{etoolbox}
\patchcmd\H@refstepcounter{\protected@edef}{\protected@xdef}{}{}
\makeatother

\newcolumntype{i}{>{\scriptsize}r}

\shorttitle{Radio Spectra of CSO-2s}
\shortauthors{De la Parra et al.}

\graphicspath{{./}{figs/}}

\crefname{equation}{Eq.}{Eqs.}
\Crefname{equation}{Equation}{Equations}
\crefname{figure}{Fig.}{Figs.}
\Crefname{figure}{Figure}{Figures}
\crefname{table}{Table}{Tables}
\Crefname{table}{Table}{Tables}
\crefname{section}{Section}{Sections}
\Crefname{section}{Section}{Sections}

\usepackage[nolist,nohyperlinks]{acronym}
\begin{acronym}
    \acro{ads}[ADS]{Astrophysics Data System}
    \acro{agn}[AGN]{Active Galactic Nucleus}
    \acroplural{agn}[AGN]{Active Galactic Nuclei}
    \acro{bologna}[Bologna]{Bologna Complete Sample of Nearby Radio Sources}
    \acro{cd}[CD]{Compact Double}
    \acro{cjf}[CJF]{Caltech Jodrell Bank flat-spectrum}
    \acro{class}[CLASS]{Cosmic Lens All-Sky Survey}
    \acro{cso}[CSO]{Compact Symmetric Object}
    \acro{css}[CSS]{Compact Steep Spectrum}
    \acro{ddrg}[DDRG]{Double-Double Radio Galaxy}
    \acroplural{ddrg}[DDRGs]{Double-Double Radio Galaxies}
    \acro{emerlin}[eMERLIN]{Enhanced Multi Element Remotely Linked Interferometer Network}
    \acro{fsrq}[FSRQ]{Flat Spectrum Radio Quasar}
    \acro{fwhm}[FWHM]{Full Width at Half Maximum}
    \acro{gps}[GPS]{Gigahertz-Peaked Spectrum}
    \acro{hfp}[HFP]{High-Frequency Peaker}
    \acro{mso}[MSO]{Medium Symmetric Object}
    \acro{ned}[NED]{NASA/IPAC Extragalactic Database}
    \acro{ovro}[OVRO]{Owens Valley Radio Observatory}
    \acro{pr}[PR]{Pearson Readhead}
    \acro{ps}[PS]{Peaked-Spectrum}
    \acro{prcj1}[PR+CJ1]{Pearson Readhead and First Caltech Jodrell Bank}
    \acro{rfc}[RFC]{Radio Fundamental Catalog}
    \acro{vips}[VIPS]{VLBA Imaging and Polarimetry Survey}
    \acro{vla}[VLA]{Very Large Array}
    \acro{vlba}[VLBA]{Very Long Baseline Array}
    \acro{vlbi}[VLBI]{Very Long Baseline Interferometry}
\end{acronym}

\begin{document}

\title{The Radio Spectra of High Luminosity Compact Symmetric Objects (CSO-2s):\\
Implications for Studies of Compact Jetted Active Galactic Nuclei}

\correspondingauthor{P. V.de la Parra}
\email{phvergara@udec.cl}

\author{P. V. de la Parra}
\affiliation{CePIA, Astronomy Department, Universidad de Concepci\'on,  Casilla 160-C, Concepci\'on, Chile}
\author{A.C.S Readhead}
\affiliation{Owens Valley Radio Observatory, California Institute of Technology,  Pasadena, CA 91125, USA}
\affiliation{Institute of Astrophysics, Foundation for Research and Technology-Hellas, GR-70013 Heraklion, Greece}
\author{T. Herbig}
\affiliation{TH Masters, LLC, 136 Chichester Rd, New Canaan, CT 06840, USA}
\author{S. Kiehlmann}
\affiliation{Institute of Astrophysics, Foundation for Research and Technology-Hellas, GR-70013 Heraklion, Greece}
\author{M.L. Lister}
\affiliation{Department of Physics and Astronomy, Purdue University, 525 Northwestern Avenue, West Lafayette, IN 47907, USA}
\author{V. Pavlidou} 
\affiliation{Institute of Astrophysics, Foundation for Research and Technology-Hellas, GR-70013 Heraklion, Greece}
\affiliation{Department of Physics and Institute of Theoretical and Computational Physics, University of Crete, 70013 Heraklion, Greece}
\author{R.A. Reeves}
\affiliation{CePIA, Astronomy Department, Universidad de Concepci\'on,  Casilla 160-C, Concepci\'on, Chile}
\author{A. Siemiginowska}
\affiliation{Center for Astrophysics|Harvard and Smithsonian, 60 Garden St., Cambridge, MA 02138, USA}
\author{A. G. Sullivan}
\affiliation{Kavli Institute for Particle Astrophysics and Cosmology, Department of Physics,
Stanford University, Stanford, CA 94305, USA}
\author{T. Surti}
\affiliation{Owens Valley Radio Observatory, California Institute of Technology,  Pasadena, CA 91125, USA}
\author{A. Synani} 
\affiliation{Department of Physics and Institute of Theoretical and Computational Physics, University of Crete, 70013 Heraklion, Greece}
\author{K. Tassis} 
\affiliation{Institute of Astrophysics, Foundation for Research and Technology-Hellas, GR-70013 Heraklion, Greece}
\affiliation{Department of Physics and Institute of Theoretical and Computational Physics, University of Crete, 70013 Heraklion, Greece}
\author{G.B. Taylor}
\affiliation{Department of Physics and Astronomy, University of New Mexico, Albuquerque, NM 87131, USA}
\author{P.N. Wilkinson}
\affiliation{Jodrell Bank Centre for Astrophysics, University of Manchester, Oxford Road, Manchester M13 9PL, UK} 
\author{M.F. Aller}
\affiliation{Department of Astronomy, University of Michigan, 323 West Hall, 1085 S. University Avenue, Ann Arbor, MI 48109, USA}
\author{R. D. Blandford}
\affiliation{Kavli Institute for Particle Astrophysics and Cosmology, Department of Physics,
Stanford University, Stanford, CA 94305, USA}
\author{N. Globus}
\affiliation{Kavli Institute for Particle Astrophysics and Cosmology, Department of Physics,
Stanford University, Stanford, CA 94305, USA}
\author{C. R. Lawrence}
\affiliation{Jet Propulsion Laboratory, California Institute of Technology, 4800 Oak Grove Drive, Pasadena, CA 91109, USA}
\author{B. Molina}
\affiliation{CePIA, Astronomy Department, Universidad de Concepci\'on,  Casilla 160-C, Concepci\'on, Chile}
\author{S. O'Neill}
\affiliation{Department of Physics, Princeton University, Jadwin Hall, Princeton,
08540, NJ, USA.}
\author{T. J. Pearson}
\affiliation{Owens Valley Radio Observatory, California Institute of Technology,  Pasadena, CA 91125, USA}

\begin{abstract}
This paper addresses, for the first time, a key aspect of the phenomenology of Compact Symmetric Objects (CSOs) -- the characteristics of their radio spectra.   We present a radio-spectrum description of a complete sample of high luminosity CSOs (CSO-2s), which shows that they exhibit the \textit{complete} range of spectral types, including flat-spectrum sources ($\alpha \ge -0.5$), steep-spectrum sources  ($\alpha < -0.5$),  and peaked-spectrum sources.  We show that there is no clear correlation between spectral type and size, but there is a correlation between the high-frequency spectral index and both object type and size.  We also show that, to avoid biasing the data and to understand the various classes of jetted-AGN involved,  the complete range of spectral types should be included in studying  the general phenomenology of compact jetted-AGN, and that complete samples must be used, selected over a wide range of frequencies. We discuss examples that demonstrate these points. We find that the high-frequency spectral indices of CSO-2s span $-1.3  <\alpha_{\rm hi} < -0.3$, and hence that radio spectral signatures cannot be used to discriminate definitively between CSO-2s, binary  galactic nuclei, and millilensed objects, unless they have $\alpha_{\rm hi} >-0.3$.

\end{abstract}


\section{Introduction}
\label{sec:intro}

Although compact jetted-AGN have been intensively studied for five decades, so that much of their phenomenology is by now well known --- see the comprehensive review of \citet{2021AandARv..29....3O} --- there are still key aspects of their phenomenology that remain to be investigated \citep{2024ApJ...961..240K,2024ApJ...961..241K,2024ApJ...961..242R}.

\begin{figure*}[hbt!]
   \centering
   \includegraphics[width=0.8\linewidth]{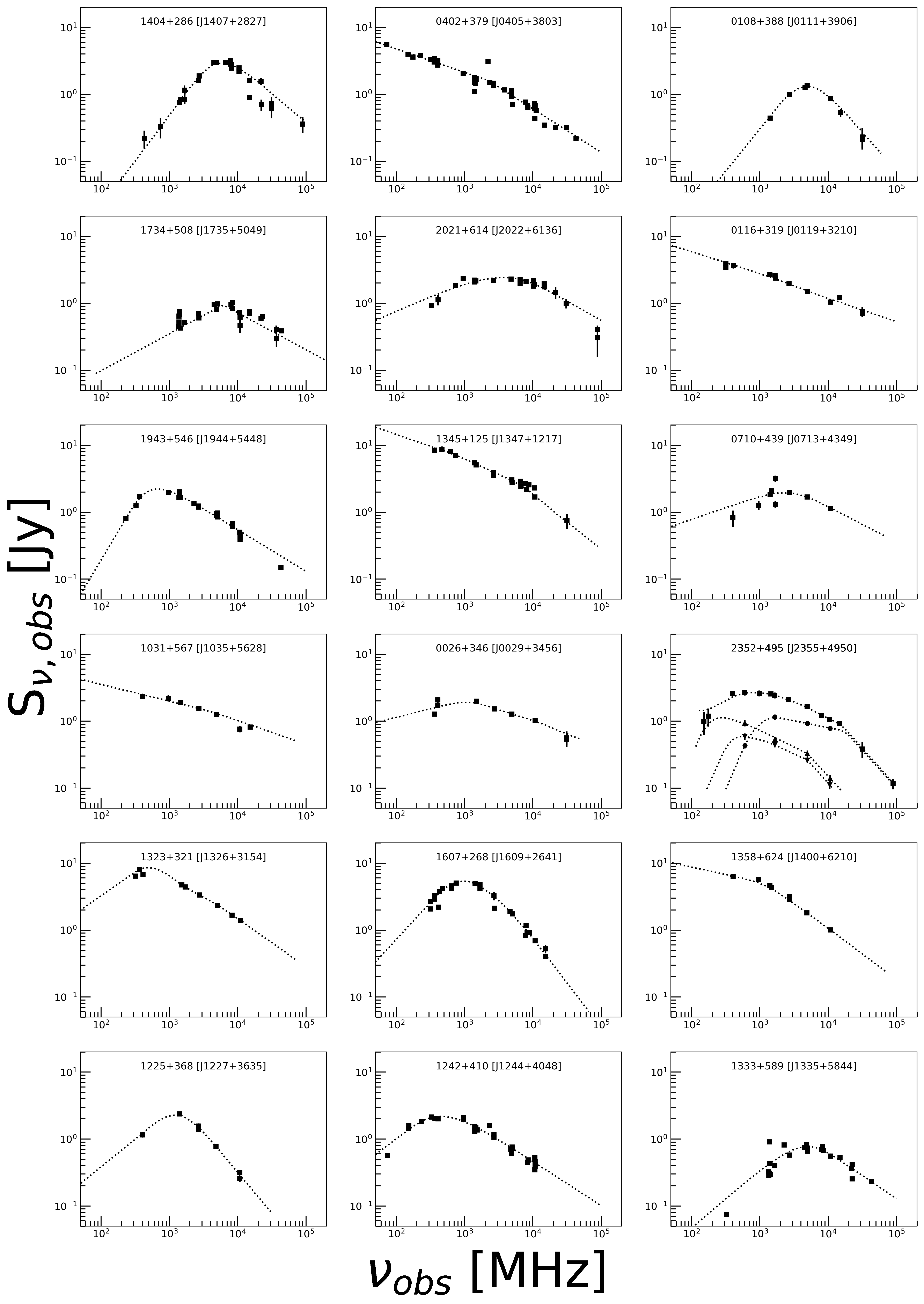}
      \caption{Radio spectra of the 17 CSO-2s with spectroscopic redshifts in the PR, CJ1 and PW complete samples arranged in  order of increasing largest physical size from left to right in successive rows.  The dotted lines show spline fits to the data.  The CSO-2 J1335+5844 is plotted last since it does not have a spectroscopic redshift, so we do not know its size. These spectra are taken from \citet{1992ApJS...81...83H},  the CATS database \citep{2005BSAO...58..118V}, and \citet{1996ApJ...460..612R}. }
         \label{plt:seventeenspectra}. 
\end{figure*}

\begin{figure*}[!t]
   \centering
   \includegraphics[width=0.8\linewidth]{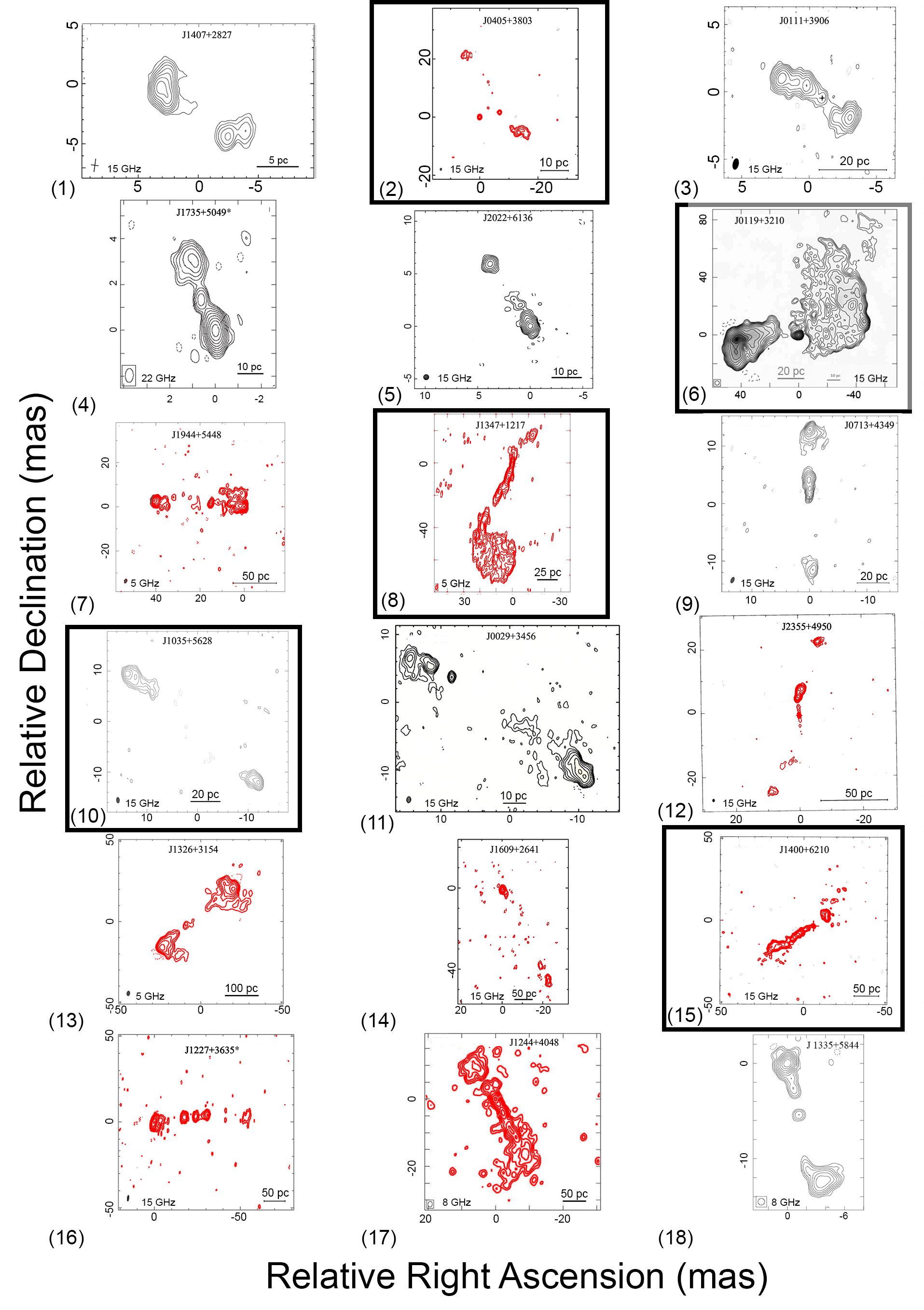}
      \caption{Hazards of  using spectral filters in selecting samples of jetted-AGN for statistically rigorous studies.  Maps of the 18 CSO-2s  in the PR, CJ1 and PW complete samples arranged in  the order of increasing size, as in  Fig. \ref{plt:seventeenspectra}, and with J1335+5844, which has no spectroscopic redshift, plotted last. The objects with red contours have steep spectra, so they would be excluded in flat-spectrum samples (see text). VLBI observations of the steep spectrum counterpart of VIPS, corresponding to the red sources, are needed for a definitive proof/disproof the sharp cutoff in CSO-2 sizes at $\sim 500$ pc (see text).  The objects in large black boxes do not have peaked spectra in the observed range, so they would be excluded in peaked-spectrum samples (see text).  The map references are as follows:1. MOJAVE, 2. \citet{2004ApJ...602..123M}, 3. \citet{1996ApJ...463...95T}, 4. \citet{2014MNRAS.438..463O}, 5. \citet{1999NewAR..43..681T}, 6. \citet{2003A&A...399..889G}, 7. \citet{1995ApJS...98....1P},
  8. \citet{1997A&A...325..943S}  , 9. \citet{2000ApJ...541..112T}, 10. \citet{2000ApJ...541..112T}, 11. \citet{2014ApJ...780..178M},  12.  \citet{1996ApJ...463...95T}, 13.  MOJAVE  , 14.  MOJAVE , 15. \citet{1996ApJ...463...95T},   16. \citet{1995ApJS...99..297X}, 17.  \citet{2004AandA...426..463O}, 18. \citet{2009AN....330..153S}.  }
         \label{plt:eighteenmaps}. 
\end{figure*}

Most studies of compact radio sources focus on the typical ``core-jet'' objects that comprise $\sim 90\%$ of the compact objects in complete samples of radio sources at centimeter wavelengths \citep{1978Natur.276..768R,1980IAUS...92..165R,1988ApJ...328..114P,1995ApJS...98....1P,1998AJ....115.1295K,2019ApJ...874...43L}.  Amongst the classes of compact radio sources that are not ``core-jets'', there is one that is of great interest both astrophysically and cosmologically that we focus on in this paper.  These are the compact symmetric objects (CSOs) \citep{1994ApJ...432L..87W,1996ApJ...460..612R}. 

\citet{2016MNRAS.459..820T} showed that there are  two fundamentally different classes of CSOs: an ``edge-dimmed'' low-luminosity class, which \citet{2024ApJ...961..240K} designated as CSO-1s, and an ``edge-brightened''  high-luminosity class, which \citet{2024ApJ...961..240K} designated as CSO-2s. In this paper we discuss only the high-luminosity CSOs, i.e., the CSO-2s.

This paper has two basic aims: (i) to reveal the basic phenomenology of the radio spectra of CSO-2s by studying a carefully-selected complete sample\footnote{A ``complete sample'' is defined to be a sample that includes all objects down to a given flux-density limit, at a specific observing frequency, and over a given area of sky \citep{1968MNRAS.139..515P,1968ApJ...151..393S,1970MNRAS.151...45L}} of CSO-2s for this purpose; and (ii) to demonstrate that only through the study of a number of such complete samples, selected over a wide range of frequencies, can biases be avoided, as is necessary for the underlying phenomenology to be revealed, thereby providing a firm foundation for the study of these objects.


The most obvious explanation of CSO-2s is that they represent the first stage in the evolution of double radio sources towards the larger Fanaroff and Riley Type II radio sources \citep{1974MNRAS.167P..31F}, to which, being edge-brightened, they are morphologically very similar.  However, as first pointed out by \citet{1994cers.conf...17R}, if they are expanding at constant speed, then the numbers indicate that strong  luminosity evolution is required such that, as they evolve from CSO-2s to FR IIs, their radio luminosit decreases by a factor $\sim 30$ while their optical narrow line luminosity increases by about the same factor. This seems a very unlikely scenario, so we reject this evolutionary scenario for CSO-2s. 

It has recently been shown that most CSO-2s are almost certainly distinct from other jetted-AGN  \citep{2024ApJ...961..241K}, and evolve through their whole life cycle in $\lesssim 5000$ yr and might result from tidal disruption of stars \citep{2024ApJ...961..242R, 2024MNRAS.528.6302S}.

 \begin{deluxetable*}{c@{\hskip 8mm}cCCCCccccc}
\tablecaption{Spectral indices and sizes of the CSO-2s in the PR, CJ1, and PW samples}
\tablehead{\colhead{B1950} & \colhead{J2000} & \colhead{Spectral index} &  \colhead{Spectral index} & \colhead{Redshift} & \colhead{Size} & Class  & PR&CJ1&PW\\
&  & ($\alpha_{2.7-5}$) & ($\alpha_{5-8}$)  && \colhead{(pc)} &  & &  }
\startdata
        B1404+286 & J1407+2827  & 0.803 & -0.037 & 0.077 & 16 & CSO-2.1  & ~ & ~ & Y  \\ 
        B0402+379 &J0405+3803 & -0.65 & -0.65 & 0.05505 & 44 & CSO-2.0 & ~ & Y & ~  \\ 
        B0108+388 &J0111+3906 &  0.471 & -0.344 & 0.668 & 56 & CSO-2.0 & Y & ~ & ~  \\ 
        B1734+508 & J1735+5049 & 0.543 & -0.035 & 0.835 & 61 & CSO-2.0 & ~ & Y & ~  \\ 
        B2021+614 & J2022+6136 & 0.012 & -0.22 & 0.227 & 105 & CSO-2.1 & Y & ~ & Y  \\ 
        B0116+319 & J0119+3210 & -0.38 & -0.387 & 0.0602 & 115 &  CSO-2.2 & ~ & ~ & Y  \\ 
        B1943+546 & J1944+5448 & -0.571 & -0.571 & 0.263 & 196 & CSO-2.0 & ~ & Y & ~  \\ 
        B1345+125 & J1347+1217 & -0.46 & -0.641 & 0.121 & 215 & CSO-2.2 & ~ & ~ & Y  \\ 
        B0710+439 & J0713+4349 & -0.269 & -0.46 & 0.518 & 217 & CSO-2.0 & Y & ~ & Y  \\ 
        B1031+567 & J1035+5628 & -0.3 & -0.328 & 0.46 & 221 & CSO-2.0 & Y & ~ & Y  \\ 
        B0026+346 & J0029+3456 & -0.324 & -0.313 & 0.517 & 259 & CSO-2.0 & ~ & ~ & Y  \\ 
        B2352+495 & J2355+4950 & -0.439 & -0.533 & 0.237 & 336 & CSO-2.2 & Y & ~ & Y  \\ 
        B1323+321 & J1326+3154 & -0.551 & -0.73 &  0.37 & 346 & CSO-2.2 & ~ & ~ & Y  \\ 
        B1607+268 & J1609+2641 & -0.973 & -1.212 & 0.473 & 362 & CSO-2.1  & ~ & ~ & Y  \\ 
        B1358+624 & J1400+6210 & -0.746 & -0.777 & 0.431 & 378 & CSO-2.2 & Y & ~ & Y  \\ 
        B1225+368 & J1227+3635 & -1.11 & -1.232 & 1.975 & 499 & CSO-2.0 & ~ & Y & Y  \\ 
        B1242+410 &J1244+4048 & -0.62 & -0.645 & 0.8135 & 529 & CSO-2.2 & ~ & Y & ~ \\ 
         B1333+589 & J1335+5844 & 0.304 & -0.206 & ~& ~ & CSO-2.0 & ~ & Y & ~ \\
\enddata
\tablecomments{The spectral indices were derived from fitting the data points observed at 2.7, 5 and 8 GHz shown in Fig. \ref{plt:seventeenspectra}, ensuring that the slope accurately represents the spectral characteristics within this frequency range. The uncertainties in the spectral indices are $\sim \pm 0.05$. The sizes and redshifts are taken from Table 3 of \citet{2024ApJ...961..240K}.}
\label{tab:zandsize}
\end{deluxetable*}

In this paper we address a basic aspect of the phenomenology of CSO-2s: their radio spectra. To do this we examine CSO-2s in three complete samples of extragalactic radio sources.  We show that about two-thirds of the objects in these samples are peaked-spectrum  sources \citep{2021AandARv..29....3O}, while about one-third of them are compact steep-spectrum sources \citep{1982MNRAS.198..843P}.  We also show that $\sim 22$\% of the sources have $\alpha_{\rm hi} >-0.5$,\footnote{We define the spectral index, $\alpha$, by $S \propto \nu^\alpha$} where $\alpha_{\rm hi}$ is the high-frequency spectral index as measured well above any peak or change in slope, and furthermore,  we find that there are roughly equal numbers of flat-spectrum $(\alpha_{\rm 5\,GHz-8\,GHz} \ge -0.5)$ and steep-spectrum $(\alpha_{\rm 5\,GHz-8\,GHz} < -0.5)$ sources in the samples we discuss.  

In view of the significant fractions of both flat and steep-spectrum sources, samples consisting of only one of these spectral types cannot be used for determining general properties of compact jetted-AGN, and similarly, samples that consist of only the peaked-spectrum sources are  incomplete for the purposes of determining the general phenomenology of these objects. Some previous searches for  CSOs, and  sample studies of jetted AGN have typically been constructed using an initial filter of flat-spectrum objects or peaked-spectrum objects only. This has led to incompleteness, for example some bright-core CSOs, like TXS 0128+554, were missed due to their having a flattish-spectrum, while others have been missed because the spectral peak lies outside the typical radio survey window of ~1 GHz - 8 GHz. 

Since the CSO classification is based solely on radio morphology, with no spectral filters applied, it provides an unbiased approach to the phenomenology of compact jetted-AGN, provided that it is applied to complete samples as defined in this paper, and also is applied to complete samples selected over a wide range of frequencies.  We give examples of samples based on radio morphology with a spectral filter applied, and/or complete samples selected over a narrow frequency range, and we point out the hazards of these approaches in each case.
\section{The Three Complete Samples}\label{sec:complete}

In a recent study of CSOs, \citet{2024ApJ...961..240K,2024ApJ...961..241K}, identified 18 bona fide CSOs (all CSO-2s), of which 17 have spectroscopic redshifts, in the following three complete radio samples:
\vskip 3pt
\noindent
(i) the Pearson-Readhead (PR) \citep{1981ApJ...248...61P,1988ApJ...328..114P}
 sample of all 65 radio sources at declinations above $\delta = 35^\circ$, with Galactic latitude  $|{b}| > 10^\circ$ and flux densities greater than 1.3\,Jy at 5\,GHz in the MPIfR/NRAO S4 and S5 surveys \citep{1978AJ.....83..451P,1981AJ.....86..854K};
 \vskip 3pt
\noindent
 (ii) 
the first Caltech-Jodrell Bank  (CJ1) sample \citep{1995ApJS...98....1P,1995ApJS...99..297X}, which extended the PR sample down to 0.7\,Jy at 5\,GHz, and added another    135 sources to the PR sample; 
\vskip 3pt
\noindent
(iii) the Peacock and Wall \citep{1981MNRAS.194..331P,1985MNRAS.216..173W} sample (PW) of 171 sources, which is complete down to $S_{\rm 2.7\,GHz} = 1.5$\,Jy over the region of sky $\delta \geq 10^\circ$ and $|b|\ge 10^\circ$,  in B1950 coordinates.  

In total there are 282 sources in these three complete samples, one of which is the starburst galaxy, M82 (3C 231), and therefore not an AGN. M82 appears in the complete sample only because its small distance and the large number of supernova remnants place its radio flux density above the PW completeness limit. Thus the parent sample of AGN that we consider below contains 281 sources.

 This combined PR+CJ1+PW AGN sample of 281 sources is unique, to the best of our knowledge, in being the only high-frequency AGN radio sample of this, or comparable, size, for which the radio structures of all of the objects are known on both large and small scales. A complete listing of this sample and its properties, together with references to maps of each object, is given by  \citet{2024ApJ...961..241K}.  It is the best available complete sample that includes significant numbers of flat- and steep-spectrum sources, as well as of gigahertz-peaked-spectrum (GPS) objects, and it is therefore ideal for investigating the complex ties between morphology and radio spectral shape in teasing out the complex phenomenology of jetted-AGN. To these complete samples, we are now engaged in observations to triple the numbers of CSOs in complete samples, and also to add complete samples selected at $\sim 1$\,GHz and at $\sim 100$\,MHz.

\section{The Radio Spectra of the CSO-2\lowercase{s} in the PR+CJ1+PW samples}\label{sec:spectra}

The spectra of the 17 CSO-2s with spectroscopic redshifts in the above three complete samples \citep{2024ApJ...961..240K} are shown in Fig.~\ref{plt:seventeenspectra}, displayed in order of increasing size, and corresponding maps are shown in Fig. \ref{plt:eighteenmaps}.

These figurea demonstrate conclusively that there is no clear relationship between spectral shape and physical size, although the larger sources do tend to peak at lower frequencies.  Of interest here are: (i) the numbers of flat and steep-spectrum sources are about equal;  (ii) approximately one third of the objects do not have peaked spectra within the observed range of frequencies; (iii) Some of  the most compact CSO-2s do not have peaked spectra within the observed range of frequencies; and (iv) some of the most extended CSO-2s have peaked spectra. 

\begin{figure}[!t]
    \centering
    \includegraphics[width=0.95\linewidth]{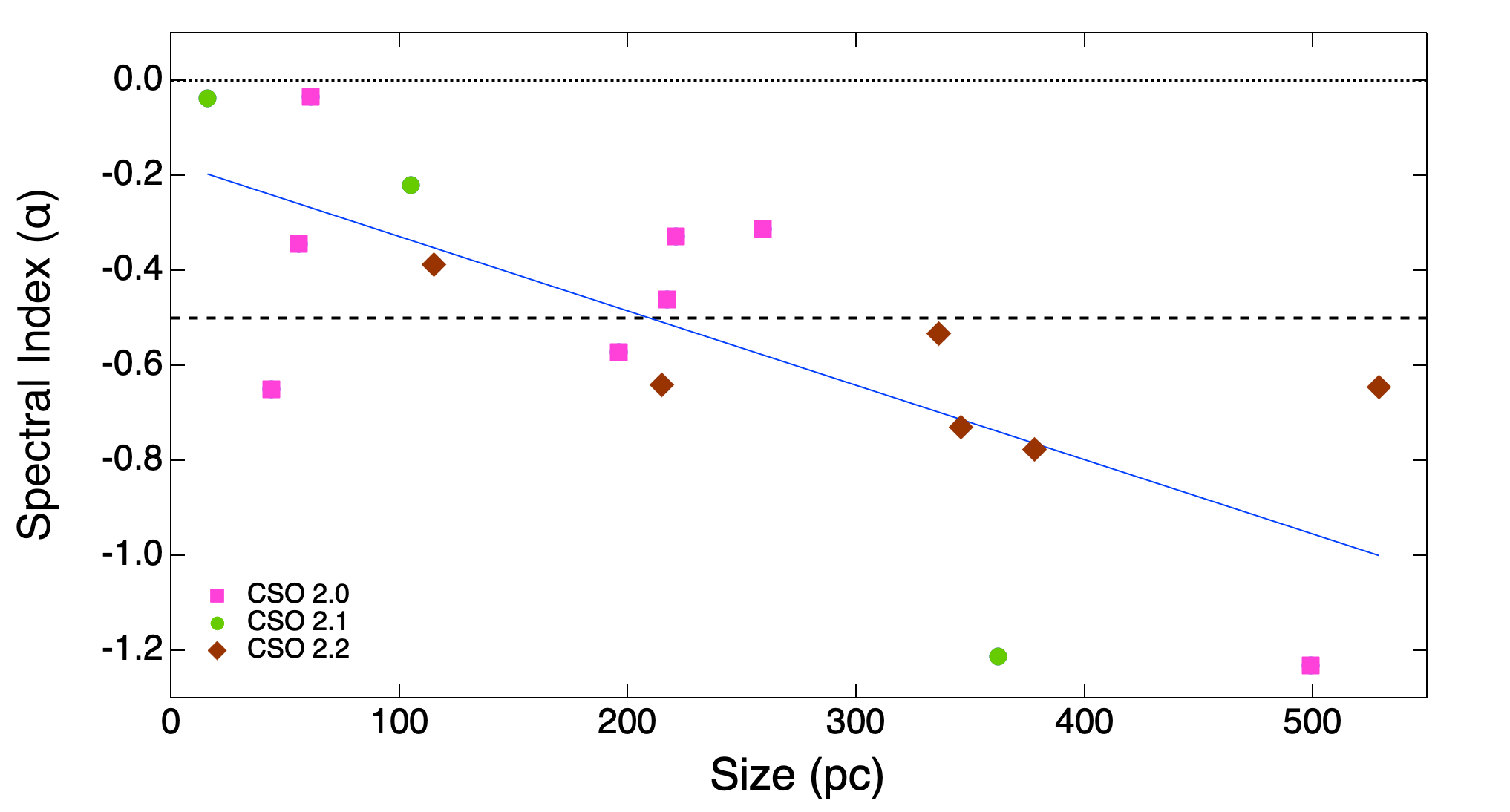}
    \caption{The 5\,GHz -- 8\,GHz spectral indices of the PR, CJ1, and PW CSOs plotted against their size. The pink squares are the CSO-2.0s, the green circles are the CSO-2.1s and the brown diamonds are the CSO-2.2s. The dotted line indicates spectral index $\alpha = 0$, and the dashed line indicates $\alpha = -0.5$. The solid line is the linear least squares fit to the data. }
    \label{plt:alpha}
\end{figure}

This clearly shows that CSOs span the whole range of spectral types, including flat-spectrum sources, steep-spectrum sources, and GPS sources. In particular, we wish to emphasize the fact that not all CSO-2s are peaked-spectrum sources within the observed range of frequencies, nor do they all have flat spectra, with spectral indices $\alpha \ge -0.5$ at cm wavelengths. In the objects that do have peaked spectra, the low-frequency cutoffs are thought to be due to synchrotron self-absorption (SSA) and/or free-free absorption (FFA). Since both SSA and FFA occur in compact objects,   it might be expected that the more compact objects would peak at higher frequencies, and indeed, as can be seen in Fig.~\ref{plt:alpha}, we find that there is a clear relationship between the overall size of these CSOs and their spectral indices  between 5\,GHz and 8\,GHz. It is important to note that, as pointed out by \citet{2021AN....342.1185R}, \textit{all} compact radio sources \textit{must} have a spectral turnover at some frequency.  We return to these points in \S \ref{sec:filter}.

The dissection of the J2355+4950 spectrum from \citet{1996ApJ...460..612R} was carried out using multi-frequency VLBI observations, and shows three SSA (or FFA) cutoffs at low frequencies and three spectral aging cutoffs at  high frequencies in various components,  is shown in Fig. \ref{plt:seventeenspectra} to illustrate the underlying typical complexity of these spectra. The redshifts, physical sizes, and spectral indices of the 17 sources in the PR+CJ1+PW complete sample  are listed in  Table \ref{tab:zandsize}, also in order of increasing size.  

The spectral indices of Fig.~\ref{plt:alpha} were all measured between 5\,GHz and 8\,GHz using the plots of Fig.~\ref{plt:seventeenspectra}.   It is immediately clear that there is a strong inverse correlation, such that larger (more positive) spectral indices correlate with smaller size. The solid line of Fig.~\ref{plt:alpha} shows a linear least squares fit to the data. The  correlation coefficient is $r = -0.76_{-0.43}^{-0.91}$, with  $p$-value $4 \times 10^{-4}$, where the lower and upper limits are 95\%  confidence limits.  

We see that between 5 GHz and 8 GHz,  9 of the 17 CSOs, i.e., $\sim 50\%$ in these complete samples, have spectral indices $\alpha < -0.5$, and thus that there are roughly equal numbers of flat-spectrum and steep-spectrum CSOs in complete samples selected at 5\,GHz.

\citet{2024ApJ...961..242R} showed that CSO-2s can be sub-divided into three morphological classes:
\vskip 6pt
\noindent
CSO~2.0 have two distinct outer lobes, making them edge-brightened, with hot spots  at their outer edges opposite the nucleus, narrow jets (if jets are visible at all), and  lobes not much wider than the hot spots. These are indicated by pink squares in \cref{plt:alpha}.

\vskip 6pt
\noindent
CSO~2.2   have two distinct outer lobes but the hot spots are not dominant or are invisible, or the lobes are well-resolved normal to the jet axis. The hotspot, or hot spots, do not need to be located at the end of the lobe opposite the core. These are indicated by brown diamonds in \cref{plt:alpha}. 

\vskip 6pt
\noindent
CSO~2.1  have some of the properties of CSO~2.0s and some of the properties of CSO~2.2s.  For example, they might look like CSO~2.0s on one side of the nucleus and CSO~2.2s on the other side, or they might have hot spots that are not located at the extremities of their envelopes.  These are indicated by green circles in \cref{plt:alpha}.

We see in  Fig.~\ref{plt:alpha} that in these complete samples CSO-2.0s in general have flatter spectra and are smaller than CSO-2.2s, and that CSO-2.1s almost span the range of CSO-2.0s and CSO-2.2s in both spectral index and size.

\begin{figure}[!b]
    \centering
    \includegraphics[width=\columnwidth]{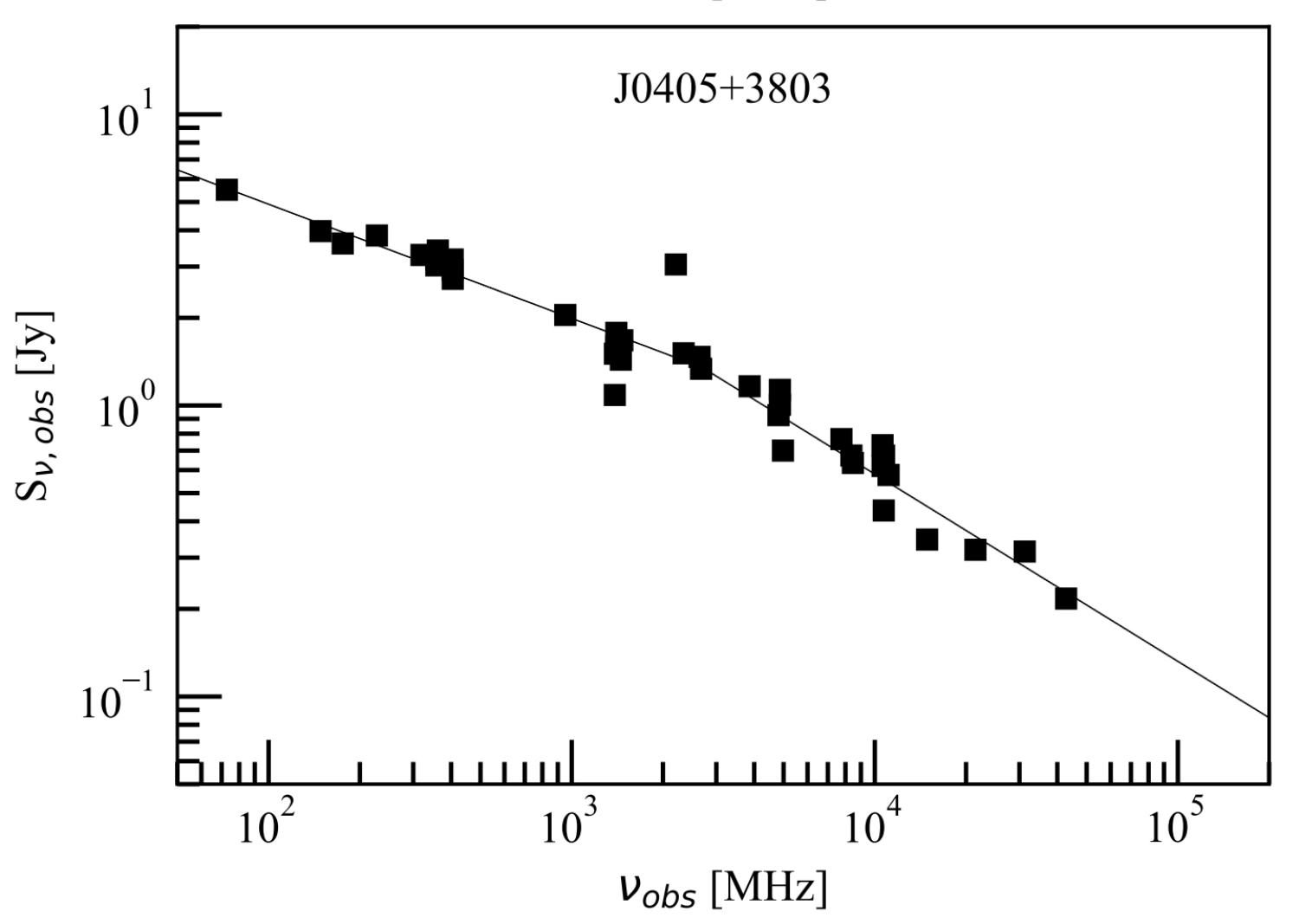}
    \caption{The spectrum of the SMBHB B0402+379 [J0405+3803]. Below 2.7 GHz the spectral index is  $\alpha =-0.35$, whereas above 2.7 GHz the spectral index is  $\alpha =-0.65$. }
    \label{plt:spec0405}
\end{figure}

The spectrum of the supermassive black hole binary (SMBHB) B0402+379 [J0405+3803] \citep{2004ApJ...602..123M,2006ApJ...646...49R,2017ApJ...843...14B} is shown in Fig.~\ref{plt:spec0405}. This is one of only three SMBHB candidates for which there is compelling evidence \citep{2004ApJ...602..123M}, the other two being OJ 287 [J0854+2006] \citep{1988ApJ...325..628S,2016ApJ...819L..37V,2021MNRAS.503.4400D} and PKS 2131--021 [J2134--0153] \citep{2022ApJ...926L..35O}.  

It is interesting to note that the spectral index of 0402+379 steepens significantly at around 2.7\,GHz, from  $\alpha \sim -0.35$ below 2.7\,GHz to $\alpha \sim -0.65$ above 2.7\,GHz.  The redshift of B0402+379 [J0405+3803] is 0.05505 \citep{2009A&A...496L...9M}, so clearly, for an object at a redshift greater than 0.05505 the spectral index above 2.7\,GHz would be steep enough to exclude it from any sample with a lower cutoff on the spectral index of $\alpha = - 0.5$.

This is a critical finding because it illustrates, in one object, the danger of applying spectral shape criteria in selecting samples for the elucidation of the phenomenology of compact radio sources. Amongst the issues discussed at the beginning of this section, we now see that spectral filters constructed using spectral index can easily 
 introduce redshift biases into the sample.  It also makes it clear that a steep high-frequency spectrum is no guarantee against an object being a supermassive black hole binary (SMBHB), or a milli-lensed galactic nucleus, because a flat-spectrum core may be dominated by the spectra of the lobes, as in this case.  In order to rule out these possibilities one needs high dynamic range maps at multiple frequencies and/or monitoring of both components of the SMBHB  to determine whether or not they show the same flux density variations.  In the case of B0402+379 [J0405+3803], differing variability over 14 years between the two compact components enabled \citet{2004ApJ...602..123M} to rule out millilensing.

\section{CSO-2 X-ray Emission}\label{sec:xray}

X-ray observations provide insight into the nature and environment of compact radio sources. The X-ray spectra provide a measurement of the total absorption towards the X-ray emitting region \citep[e.g.][]{2009AN....330..264S, 2017ApJ...849...34O}.
\cite{2019ApJ...871...71S} show that the most compact CSOs ($<60$pc, i.e. those of approximately the size of B0108+388),  have high intrinsic absorption and may reside in dense environments. Sixteen of the sources in Table~\ref{tab:zandsize} have X-ray data (those without are B1734+508 and  B1225+368). In comparing the radio spectral index with the HI absorption column for the objects in our sample we find no clear trend. 

X-rays together with a broad-band modeling of the spectral energy distribution (SED) can provide constraints on emission models and discussions of the source nature. Recently, \cite{2024ApJ...966..201K} investigated a model with  X-ray emission originating from the compact radio lobes expanding  within the dusty-molecular medium. Here, the compact radio structures would be fully embedded in a dense nuclear medium. They argue that the observed X-ray radiation in this case is due to the inverse-Compton emission of the compact radio lobes with the X-ray spectral features being due to  reflection of the primary continuum off the surrounding dust. They also discuss a correlation between the spectral index of the electron distribution in the lobes and the X-ray absorption column. We note that the two sources in their studies,  B1404+286 and B2021+614, both have a relatively flat spectral index in Table~\ref{tab:zandsize}, and show spectral steepening at the higher frequencies covered by their sub-mm observations. 
The sub-mm data were critical for constraining the electron distribution and other parameters in this model. A complete sample of compact radio sources with a good spectral energy distribution (SED) coverage could provide a test of the  origin of the broad-band radiation. Note that the lobe radiation becomes less dominant in larger sources that have evolved beyond the central dusty medium.

\section{Pitfalls of Spectrally Incomplete Radio Samples}\label{sec:filter}

 In this section we discuss some of the issues that can arise when one sets out to investigate the fundamental phenomenology of compact jetted-AGN using spectrally incomplete samples. We give below two examples. In the first only flat-spectrum sources are considered, in the second a limited frequency range is used  to define a complete sample. We also point out that peaked-spectrum sources have the same hazards as, for example, flat-spectrum samples.  The result of these spectral filters is that significant fractions of the populations under study are likely to  be excluded.

\subsection{Flat-spectrum $(\alpha \geq -0.5)$ and Peaked-spectrum Samples }\label{sec:flat}

 Early very long baseline (VLBI) surveys of compact radio sources focused on complete samples \citep{1981ApJ...248...61P,1995ApJS...98....1P}. Thereafter, it was correctly thought that  searches for supermassive black hole binaries (SMBHBs) and gravitational milli-lenses would be more efficient were they to focus on flat-spectrum sources,  with spectral indices $\alpha \ge -0.5$, because these automatically select for compact, core-dominated, objects. This approach was adopted in the Caltech-Jodrell flat-spectrum survey (CJF) \citep{1994ApJS...95..345T}, the CLASS survey \citep{2003MNRAS.341....1M,2003MNRAS.341...13B}, and the VLBA imaging and polarimetry survey \citep{2007ApJ...658..203H} (VIPS).  

It is clear from Fig.~\ref{plt:alpha} that  about half of the  CSOs will be missed in any sample selected at centimeter wavelengths that excludes steep-spectrum sources with spectral indices $\alpha < -0.5$. Therefore the CJF, CLASS,  and VIPS samples have likely missed about half of the CSOs that would be found in complete samples down to their flux density limits. It is therefore clearly essential to include the steep-spectrum sources in any studies of CSOs that one wishes to  use in  statistical or phenomenological studies of the CSO-2 class. 

We now explore one consequence for statistical studies of selecting a flat-spectrum sample in more detail.  In the VIPS survey, \citet{2007ApJ...658..203H} and \citet{2016MNRAS.459..820T} defined a sample of 1127 flat-spectrum sources ($\alpha \ge -0.5$) from the CLASS survey \citep{2003MNRAS.341....1M,2003MNRAS.341...13B}  and observed all 1127 sources at 5\,GHz on the VLBA. 
They subsequently did follow-up studies on CSO and SMBHB candidates at multiple frequencies, and discovered 20 new CSOs. So this program was successful for the purpose of efficient discovery of CSOs, but it  was less helpful for the purposes of understanding the phenomenology of CSOs. For example, there are 11 VIPS CSO-2s that are not in the PR+CJ1+PW sample, and all but one of these has size less than 500\,pc. For the sake of argument, if we were to ignore the possible biases, and  add these VIPS CSO-2s to the 17 CSO-2s in the PR+CJ1+PW sample, which also has only one CSO-2 of size $>500$ pc, then there would be 26 out of 28 CSO-2s in the resulting sample with size $< 500$ pc.  The binomial probability of this 26:2 ratio is $P=1.4\times 10^{-6}$, i.e., a $4.7 \sigma$ result, which would amount to a compelling confirmation of the $\sim 500$\,pc size cutoff in CSO-2s found by \citet{2024ApJ...961..241K}.  However, from Fig.~\ref{plt:alpha} and Table~\ref{tab:zandsize}, we see that roughly half the CSO-2s in complete samples have steep high-frequency spectra, and so they would have been excluded by the VIPS steep-spectrum cutoff. We also see from Fig.~\ref{plt:alpha} that the size of the CSO-2s increases with steepening radio spectrum. 

We have applied for observing time for the steep-spectrum counterpart of the VIPS sample. We refer to this new complete sample, composed of both the flat and the steep-spectrum sources in the VIPS footprint with $S_{\rm 8\,GHz} \ge 85$\,mJy, as the \textit{Extended} VIPS (EVIPS) sample. The EVIPS sample contains 2121 sources. It is to be expected that the EVIPS sample will contain approximately twice as many CSOs as the original VIPS sample.
It is entirely possible that a significant fraction of the EVIPs steep-spectrum CSOs will have sizes greater than 500 pc. If, for example, all of the expected 11 sources have  size greater than 500\,pc, then there would be 26~sources with size less than 500\,pc and 12~sources with size greater than 500\,pc, and the probability rises to 0.01, i.e., only a  2.3$\sigma$ result --- down from the $P=1.7 \times 10^{-4}$, $(3.6\sigma)$ result of \citet{2024ApJ...961..241K}.  This would, in our view, significantly dilute the evidence for a sharp cutoff in sizes. 

The size of the EVIPS sample (2121 sources) may be compared to the complete sample of 281 sources discussed by \citet{2024ApJ...961..241K}. If the EVIPS steep-spectrum CSO 2s show the same size distribution as the present complete sample, and the same size distribution as the VIPS flat-spectrum sources, then the level of significance will increase to $p$-value $\sim 2\times 10^{-9}$ ($\sim 6 \sigma$), leaving virtually no room for  doubt about  the size cutoff observed by \citet{2024ApJ...961..241K}, and hence of the \textit{distinctive} nature of the majority of CSO-2s.

It is clear from the discussion in this section that samples based on peaked-spectrum sources will suffer the same fate as do flat-spectrum samples or steep-spectrum samples, because they will miss a significant fraction of the populations under study, thereby potentially biasing the results.

 We have included the above flat-spectrum sample example for three reasons: (i) to demonstrate that the test we are proposing could disprove or prove the sharp size cutoff and hence the distinctive nature of most CSO-2s;  (ii) to emphasize the point that the inclusion of a radio spectrum filter can in principle dramatically bias the basic phenomenology under study; and (iii) to make a compelling case for follow-up of the sharp cutoff in the CSO-2 size distribution found by \citet{2024ApJ...961..241K}, which we turn to next.

 \subsubsection{Verification or Disproof of the CSO-2 Size Cutoff}\label{sec:verification}
 
The fact that the apparent size cutoff in CSO-2s found by \citet{2024ApJ...961..241K} is  only a factor $\sim 2$ below the upper cutoff at 1\,kpc that defines the class is no coincidence. The upper size limit of 1\,kpc was set by \citet{1996ApJ...460..612R} based on the first five CSOs  to be discovered, which are all CSO-2s: 0108+388 (56\,pc), 0404+768 (44\,pc), 0710+439 (217\,pc), 1358+624 (378\,pc), and 2352+495 (336\,pc). 

It appeared at that time, therefore,  that there was an upper cutoff in the sizes of CSO-2s at around 400\,pc. However, since the numbers were based on only five objects, it was decided by those authors to set the upper size limit criterion for classification as a CSO  at 1\,kpc in order to allow for some CSOs that might be larger than those that had been discovered up to that time.  \citet{2024ApJ...961..241K} have now found a statistically significant reduction,  at the $p$-value $=1.7 \times 10^{-4}$ ($3.6\sigma$),  in the numbers of CSO-2s in the 500\,pc--1\,kpc size range. In the three complete samples described in \S \ref{sec:complete} there are 17 CSO-2s, i.e., an increase in the size of the sample by more than a factor of 3 over the five CSO-2s studied by \citet{1996ApJ...460..612R}, and these 17 CSO-2s show the same size distribution as the first five listed above. The careful statistical evaluation of the size distribution of these 17 CSOs, using both the Kolmogorov-Smirnoff test and the binomial test, returned the above numbers. It is clearly important to test whether the observed size cutoff is real or due to the small size of this sample. By more than trebling the sample size, to $\sim61$ CSOs, it should be possible to settle this question once and for all. A VLBA proposal to do just this is pending. Verification of the size cutoff would provide compelling proof that the vast majority of CSO-2s are distinct from the other classes of jetted-AGN. An upper cutoff in size implies an upper cutoff in energy, raising the fascinating possibility that most CSO-2s are formed by the capture and tidal disruption of a single star by a latent SMBH \citep{1994cers.conf...17R,2012ApJ...760...77A,2024ApJ...961..242R,2024MNRAS.528.6302S}.

\subsection{Samples Selected at Only One Frequency}\label{sec:onefreq}

Inspection of Fig.~\ref{plt:seventeenspectra} shows that half of the CSO-2s in this complete sample would be missed, were we to use a flux-density-limited sample of sources with $S_{\rm 100\,MHz}\ge 0.7$\,Jy (i.e., the same flux-density cutoff that we have used at 5\,GHz).  It is, therefore, critically important to cover a wide range of frequencies in which the sample is complete above a specified flux density limit  when exploring the phenomenology of compact jetted-AGN.  For this reason we are now observing samples that are complete at 178\,MHz and 966\,MHz to match those used by \citet{2024ApJ...961..240K} at 2.7\,GHz and 5\,GHz.

\begin{figure}[!b]
    \centering
    \includegraphics[width=0.95\linewidth]{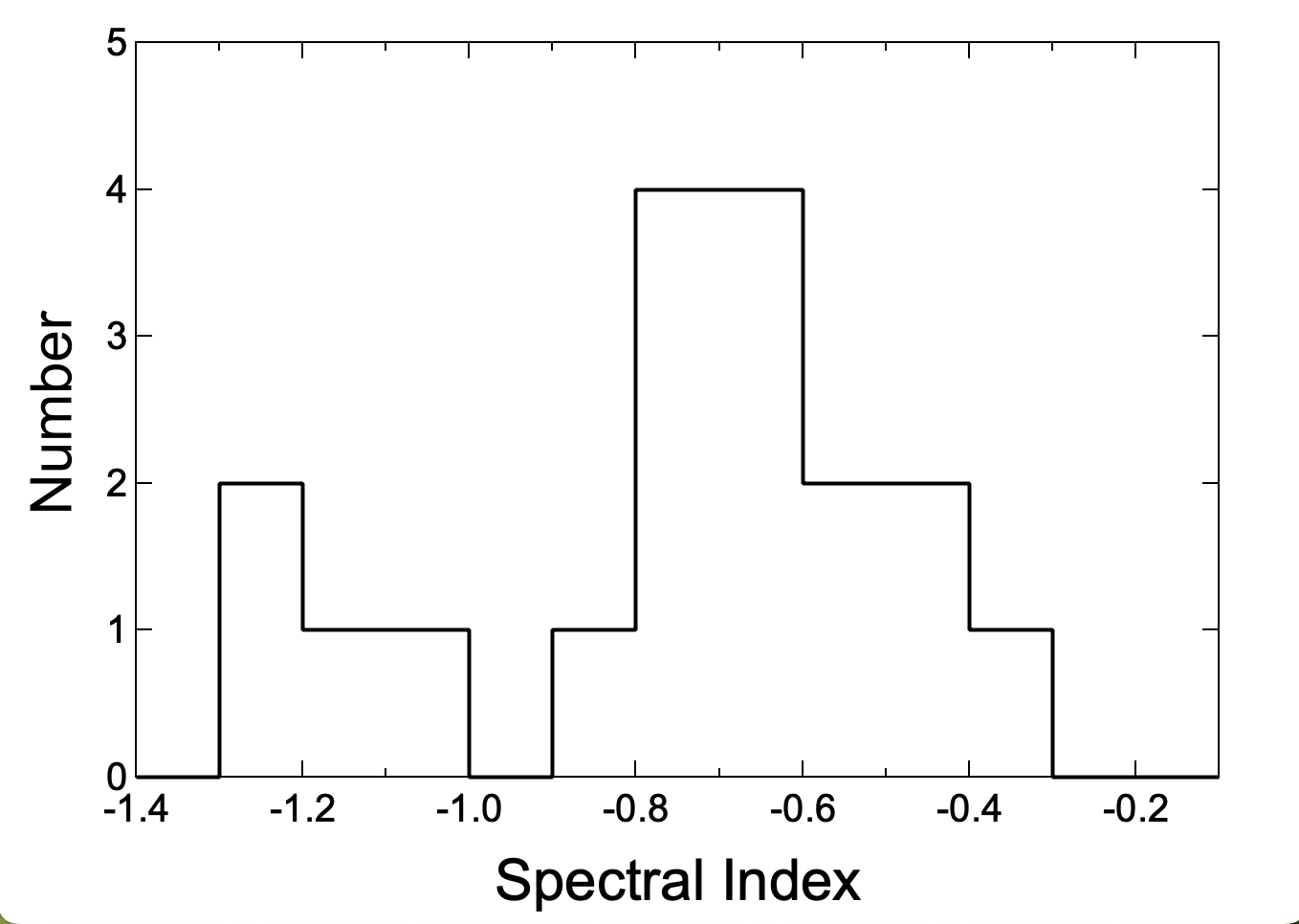}
    \caption{The high-frequency spectral index, $\alpha_{\rm hi}$, distribution of the PR, CJ1, and PW CSO-2s. This includes the CSO-2 1335+5844 ($\alpha_{\rm hi}=-0.70$), for which there is no spectroscopic redshift.}
    \label{plt:alphahi}
\end{figure}

\section{Discriminating between CSO\lowercase{s}, Binary Nuclei, and  Millilensed objects}\label{sec:steep}

\subsection{Radio Spectrum Discrimination}\label{sec:radspec}

An important aspect of compact jetted-AGN studies is that of searching for Binary Nuclei (BN), and millilenses in the mass range $10^6\,{\rm M}_\odot$--$3 \times 10^9\,{\rm M}_\odot$ \citep{2001PhRvL..86..584W,2021MNRAS.507L...6C}.   CSO-2s are the most serious contaminating objects in such searches. 

One potential method of discriminating CSO-2s from BN and millilenses is the high-frequency spectral index, i.e. the spectral index at the top end of the observed spectrum, above the region affected by any peak or break in the spectrum. These high-frequency spectral slopes were estimated by eye in straightforward cases, and using least squares fits if the fit by eye was not obvious.  Note that these spectral indices  are not the same as the $\alpha_{5-8}$ values of Table \ref{tab:zandsize}. The high-frequency spectral index distribution of the 18 CSO-2s in the complete samples discussed in \S~\ref{sec:complete} is shown in Fig.~\ref{plt:alphahi}.  We see here that the high-frequency spectral indices of CSO-2s range from $-1.3$ to $-0.3$.  Therefore, when using the spectral index as a filter for eliminating CSO-2s in BN and/or millilense searches, the spectral index may possibly provide a definitive filter for objects with $\alpha > -0.3$, but it clearly does not provide a definitive filter for objects with $\alpha \leq -0.3$.

The number of strong millilens and/or SMBHB candidates is likely to be small \citep{2001PhRvL..86..584W}, and the astrophysical importance of finding an SMBHB or millilens is sufficiently great that observing time at both optical and radio frequencies is likely to be awarded for any compelling candidates.  Thus the considerations discussed in this section should serve only as a guide as to what might be happening in CSO-2s that could  appear in searches for such objects, and should not be used to exclude potential strong SMBHB or millilensing candidates, either of which would be an important discovery.

\subsection{Radio Light Curve Discrimination}\label{sec:radligth}

   \citet{2004ApJ...602..123M} studied the variability  B0402+379 [J0405+3803] over 14~years with VLBI.  They were able to discriminate between an SMBHB and a millilens because the two components that they were interested in were the two core components, which showed appreciably different variability over the 14-year timescale.    In the results of \citet{2004ApJ...602..123M} it can be seen that the two lobes show very little  variability, and it would almost certainly not be possible to discriminate between these and a slowly-varying millilensed system. The lobes of most CSOs vary only slowly, if at all, so this degeneracy between CSOs/SMBHBs/Milli-lenses can be very hard to break without VLBI observations extending over at least a decade, and preferably two decades, in such cases.

\section{Discussion}\label{sec:discussion}

We have shown that the radio spectra of complete samples of CSO-2 objects cover the whole range of radio spectral types, including flat-spectrum, steep-spectrum, and peaked-spectrum sources.  We have also shown that the CSO-2 sources that are not peaked, amounting to $\sim$ one third of them, span the range of sizes, and are not predominantly to be found at the large end of the size range, as might have been expected.  We have also shown that the 5--8 GHz spectral indices of CSO-2s show a strong correlation with size.  We conclude that it is extremely important in  studies of the phenomenology of compact jetted-AGN not to introduce spectral filters of any type into the source selection as this can bias the outcome.

It is clearly important to test whether the sharp cutoff in the CSO-2 size distribution discovered by \citet{2024ApJ...961..241K}, is real or due to the small sample, so this should be vigorously pursued.  We propose to do this by more than trebling the sample size, from 17 to $\sim61$ CSO-2s.   Verification of the size cutoff would provide compelling proof that the vast majority of CSO-2s are distinct from the other classes of jetted-AGN, and if proven it could prove a watershed for AGN studies.   It would put a spotlight on the population of compact radio sources in which the observed emission is not strongly beamed towards the observer, and in the case of the smallest CSO-2s we would see significant structural evolution on a timescale of years. Such studies would enable, for the first time, \textit{direct} testing of both the theory and the simulations of evolving relativistic jets in AGN, which remain, otherwise, virtually untestable in the critical phases of the initial launching of the jet, and the life-cycle evolution of double radio sources.

\acknowledgments

G. Taylor acknowledges support for this research by NASA through Fermi Guest investigator grant 80NSSC22K1939.  N. Globus acknowledges support by the Simons Foundation (00001470). R.R. and B.M. and P.V.d.l.P. acknowledge support from ANID Basal AFB-170002, from Núcleo Milenio TITANs (NCN2023\_002), and CATA BASAL FB210003. P.V.d.l.P. also acknowledges support by the National Agency for Research and Development
(ANID) / Scholarship Program / Doctorado Nacional/2023--21232103. In addition, gratefully acknowledge the support provided by the National Science Foundation (NSF) Grant AST-2407603.
This paper depended on a very large amount of \ac{vlbi} data, almost all of which was taken with the Very Long Baseline Array. The National Radio Astronomy Observatory is a facility of the National Science Foundation operated under cooperative agreement by Associated Universities, Inc.  Parts of the work were carried out at the Jet Propulsion Laboratory, California Institute of Technology, under a contract with the National Aeronautics and Space Administration.

\clearpage

\bibliography{main}{}

\begin{thebibliography}{}
\expandafter\ifx\csname natexlab\endcsname\relax\def\natexlab#1{#1}\fi
\providecommand{\url}[1]{\href{#1}{#1}}
\providecommand{\dodoi}[1]{doi:~\href{http://doi.org/#1}{\nolinkurl{#1}}}
\providecommand{\doeprint}[1]{\href{http://ascl.net/#1}{\nolinkurl{http://ascl.net/#1}}}
\providecommand{\doarXiv}[1]{\href{https://arxiv.org/abs/#1}{\nolinkurl{https://arxiv.org/abs/#1}}}

\bibitem[{{An} \& {Baan}(2012)}]{2012ApJ...760...77A}
{An}, T., \& {Baan}, W.~A. 2012, \apj, 760, 77, \dodoi{10.1088/0004-637X/760/1/77}

\bibitem[{{Bansal} {et~al.}(2017){Bansal}, {Taylor}, {Peck}, {Zavala}, \& {Romani}}]{2017ApJ...843...14B}
{Bansal}, K., {Taylor}, G.~B., {Peck}, A.~B., {Zavala}, R.~T., \& {Romani}, R.~W. 2017, \apj, 843, 14, \dodoi{10.3847/1538-4357/aa74e1}

\bibitem[{{Browne} {et~al.}(2003){Browne}, {Wilkinson}, {Jackson}, {Myers}, {Fassnacht}, {Koopmans}, {Marlow}, {Norbury}, {Rusin}, {Sykes}, {Biggs}, {Blandford}, {de Bruyn}, {Chae}, {Helbig}, {King}, {McKean}, {Pearson}, {Phillips}, {Readhead}, {Xanthopoulos}, \& {York}}]{2003MNRAS.341...13B}
{Browne}, I.~W.~A., {Wilkinson}, P.~N., {Jackson}, N.~J.~F., {et~al.} 2003, \mnras, 341, 13, \dodoi{10.1046/j.1365-8711.2003.06257.x}

\bibitem[{{Casadio} {et~al.}(2021){Casadio}, {Blinov}, {Readhead}, {Browne}, {Wilkinson}, {Hovatta}, {Mandarakas}, {Pavlidou}, {Tassis}, {Vedantham}, {Zensus}, {Diamantopoulos}, {Dolapsaki}, {Gkimisi}, {Kalaitzidakis}, {Mastorakis}, {Nikolaou}, {Ntormousi}, {Pelgrims}, \& {Psarras}}]{2021MNRAS.507L...6C}
{Casadio}, C., {Blinov}, D., {Readhead}, A.~C.~S., {et~al.} 2021, \mnras, 507, L6, \dodoi{10.1093/mnrasl/slab082}

\bibitem[{{Dey} {et~al.}(2021){Dey}, {Valtonen}, {Gopakumar}, {Lico}, {G{\'o}mez}, {Susobhanan}, {Komossa}, \& {Pihajoki}}]{2021MNRAS.503.4400D}
{Dey}, L., {Valtonen}, M.~J., {Gopakumar}, A., {et~al.} 2021, \mnras, 503, 4400, \dodoi{10.1093/mnras/stab730}

\bibitem[{{Fanaroff} \& {Riley}(1974)}]{1974MNRAS.167P..31F}
{Fanaroff}, B.~L., \& {Riley}, J.~M. 1974, \mnras, 167, 31P, \dodoi{10.1093/mnras/167.1.31P}

\bibitem[{{Giroletti} {et~al.}(2003){Giroletti}, {Giovannini}, {Taylor}, {Conway}, {Lara}, \& {Venturi}}]{2003A&A...399..889G}
{Giroletti}, M., {Giovannini}, G., {Taylor}, G.~B., {et~al.} 2003, \aap, 399, 889, \dodoi{10.1051/0004-6361:20021821}

\bibitem[{{Helmboldt} {et~al.}(2007){Helmboldt}, {Taylor}, {Tremblay}, {Fassnacht}, {Walker}, {Myers}, {Sjouwerman}, {Pearson}, {Readhead}, {Weintraub}, {Gehrels}, {Romani}, {Healey}, {Michelson}, {Blandford}, \& {Cotter}}]{2007ApJ...658..203H}
{Helmboldt}, J.~F., {Taylor}, G.~B., {Tremblay}, S., {et~al.} 2007, \apj, 658, 203, \dodoi{10.1086/511005}

\bibitem[{{Herbig} \& {Readhead}(1992)}]{1992ApJS...81...83H}
{Herbig}, T., \& {Readhead}, A. C.~S. 1992, \apjs, 81, 83, \dodoi{10.1086/191688}

\bibitem[{{Kellermann} {et~al.}(1998){Kellermann}, {Vermeulen}, {Zensus}, \& {Cohen}}]{1998AJ....115.1295K}
{Kellermann}, K.~I., {Vermeulen}, R.~C., {Zensus}, J.~A., \& {Cohen}, M.~H. 1998, \aj, 115, 1295, \dodoi{10.1086/300308}

\bibitem[{{Kiehlmann} {et~al.}(2024{\natexlab{a}}){Kiehlmann}, {Lister}, {Readhead}, {Liodakis}, {O'Neill}, {Pearson}, {Sheldahl}, {Siemiginowska}, {Tassis}, {Taylor}, \& {Wilkinson}}]{2024ApJ...961..240K}
{Kiehlmann}, S., {Lister}, M.~L., {Readhead}, A.~C.~S., {et~al.} 2024{\natexlab{a}}, \apj, 961, 240, \dodoi{10.3847/1538-4357/ad0c56}

\bibitem[{{Kiehlmann} {et~al.}(2024{\natexlab{b}}){Kiehlmann}, {Readhead}, {O'Neill}, {Wilkinson}, {Lister}, {Liodakis}, {Bruzewski}, {Pavlidou}, {Pearson}, {Sheldahl}, {Siemiginowska}, {Tassis}, \& {Taylor}}]{2024ApJ...961..241K}
{Kiehlmann}, S., {Readhead}, A.~C.~S., {O'Neill}, S., {et~al.} 2024{\natexlab{b}}, \apj, 961, 241, \dodoi{10.3847/1538-4357/ad0cc2}

\bibitem[{{Kr{\'o}l} {et~al.}(2024){Kr{\'o}l}, {Sobolewska}, {Stawarz}, {Siemiginowska}, {Migliori}, {Principe}, \& {Gurwell}}]{2024ApJ...966..201K}
{Kr{\'o}l}, D.~{\L}., {Sobolewska}, M., {Stawarz}, {\L}., {et~al.} 2024, \apj, 966, 201, \dodoi{10.3847/1538-4357/ad3632}

\bibitem[{{Kuehr} {et~al.}(1981){Kuehr}, {Pauliny-Toth}, {Witzel}, \& {Schmidt}}]{1981AJ.....86..854K}
{Kuehr}, H., {Pauliny-Toth}, I.~I.~K., {Witzel}, A., \& {Schmidt}, J. 1981, \aj, 86, 854, \dodoi{10.1086/112957}

\bibitem[{{Lister} {et~al.}(2019){Lister}, {Homan}, {Hovatta}, {Kellermann}, {Kiehlmann}, {Kovalev}, {Max-Moerbeck}, {Pushkarev}, {Readhead}, {Ros}, \& {Savolainen}}]{2019ApJ...874...43L}
{Lister}, M.~L., {Homan}, D.~C., {Hovatta}, T., {et~al.} 2019, \apj, 874, 43, \dodoi{10.3847/1538-4357/ab08ee}

\bibitem[{{Longair} \& {Scheuer}(1970)}]{1970MNRAS.151...45L}
{Longair}, M.~S., \& {Scheuer}, P.~A.~G. 1970, \mnras, 151, 45, \dodoi{10.1093/mnras/151.1.45}

\bibitem[{{Maness} {et~al.}(2004){Maness}, {Taylor}, {Zavala}, {Peck}, \& {Pollack}}]{2004ApJ...602..123M}
{Maness}, H.~L., {Taylor}, G.~B., {Zavala}, R.~T., {Peck}, A.~B., \& {Pollack}, L.~K. 2004, \apj, 602, 123, \dodoi{10.1086/380919}

\bibitem[{{Marr} {et~al.}(2014){Marr}, {Perry}, {Read}, {Taylor}, \& {Morris}}]{2014ApJ...780..178M}
{Marr}, J.~M., {Perry}, T.~M., {Read}, J., {Taylor}, G.~B., \& {Morris}, A.~O. 2014, \apj, 780, 178, \dodoi{10.1088/0004-637X/780/2/178}

\bibitem[{{Morganti} {et~al.}(2009){Morganti}, {Emonts}, \& {Oosterloo}}]{2009A&A...496L...9M}
{Morganti}, R., {Emonts}, B., \& {Oosterloo}, T. 2009, \aap, 496, L9, \dodoi{10.1051/0004-6361/200911705}

\bibitem[{{Myers} {et~al.}(2003){Myers}, {Jackson}, {Browne}, {de Bruyn}, {Pearson}, {Readhead}, {Wilkinson}, {Biggs}, {Blandford}, {Fassnacht}, {Koopmans}, {Marlow}, {McKean}, {Norbury}, {Phillips}, {Rusin}, {Shepherd}, \& {Sykes}}]{2003MNRAS.341....1M}
{Myers}, S.~T., {Jackson}, N.~J., {Browne}, I.~W.~A., {et~al.} 2003, \mnras, 341, 1, \dodoi{10.1046/j.1365-8711.2003.06256.x}

\bibitem[{{O'Dea} \& {Saikia}(2021)}]{2021AandARv..29....3O}
{O'Dea}, C.~P., \& {Saikia}, D.~J. 2021, \aapr, 29, 3, \dodoi{10.1007/s00159-021-00131-w}

\bibitem[{{O'Neill} {et~al.}(2022){O'Neill}, {Kiehlmann}, {Readhead}, {Aller}, {Blandford}, {Liodakis}, {Lister}, {Mr{\'o}z}, {O'Dea}, {Pearson}, {Ravi}, {Vallisneri}, {Cleary}, {Graham}, {Grainge}, {Hodges}, {Hovatta}, {L{\"a}hteenm{\"a}ki}, {Lamb}, {Lazio}, {Max-Moerbeck}, {Pavlidou}, {Prince}, {Reeves}, {Tornikoski}, {Vergara de la Parra}, \& {Zensus}}]{2022ApJ...926L..35O}
{O'Neill}, S., {Kiehlmann}, S., {Readhead}, A.~C.~S., {et~al.} 2022, \apjl, 926, L35, \dodoi{10.3847/2041-8213/ac504b}

\bibitem[{{Orienti} \& {Dallacasa}(2014)}]{2014MNRAS.438..463O}
{Orienti}, M., \& {Dallacasa}, D. 2014, \mnras, 438, 463, \dodoi{10.1093/mnras/stt2217}

\bibitem[{{Orienti} {et~al.}(2004){Orienti}, {Dallacasa}, {Fanti}, {Fanti}, {Tinti}, \& {Stanghellini}}]{2004AandA...426..463O}
{Orienti}, M., {Dallacasa}, D., {Fanti}, C., {et~al.} 2004, \aap, 426, 463, \dodoi{10.1051/0004-6361:20041204}

\bibitem[{{Ostorero} {et~al.}(2017){Ostorero}, {Morganti}, {Diaferio}, {Siemiginowska}, {Stawarz}, {Moderski}, \& {Labiano}}]{2017ApJ...849...34O}
{Ostorero}, L., {Morganti}, R., {Diaferio}, A., {et~al.} 2017, \apj, 849, 34, \dodoi{10.3847/1538-4357/aa8ef6}

\bibitem[{{Pauliny-Toth} {et~al.}(1978){Pauliny-Toth}, {Witzel}, {Preuss}, {K{\"u}hr}, {Kellermann}, {Fomalont}, \& {Davis}}]{1978AJ.....83..451P}
{Pauliny-Toth}, I.~I.~K., {Witzel}, A., {Preuss}, E., {et~al.} 1978, \aj, 83, 451, \dodoi{10.1086/112223}

\bibitem[{{Peacock} \& {Wall}(1981)}]{1981MNRAS.194..331P}
{Peacock}, J.~A., \& {Wall}, J.~V. 1981, \mnras, 194, 331, \dodoi{10.1093/mnras/194.2.331}

\bibitem[{{Peacock} \& {Wall}(1982)}]{1982MNRAS.198..843P}
---. 1982, \mnras, 198, 843, \dodoi{10.1093/mnras/198.3.843}

\bibitem[{{Pearson} \& {Readhead}(1981)}]{1981ApJ...248...61P}
{Pearson}, T.~J., \& {Readhead}, A.~C.~S. 1981, \apj, 248, 61, \dodoi{10.1086/159130}

\bibitem[{{Pearson} \& {Readhead}(1988)}]{1988ApJ...328..114P}
---. 1988, \apj, 328, 114, \dodoi{10.1086/166274}

\bibitem[{{Polatidis} {et~al.}(1995){Polatidis}, {Wilkinson}, {Xu}, {Readhead}, {Pearson}, {Taylor}, \& {Vermeulen}}]{1995ApJS...98....1P}
{Polatidis}, A.~G., {Wilkinson}, P.~N., {Xu}, W., {et~al.} 1995, \apjs, 98, 1, \dodoi{10.1086/192152}

\bibitem[{{Pooley} \& {Ryle}(1968)}]{1968MNRAS.139..515P}
{Pooley}, G.~G., \& {Ryle}, M. 1968, \mnras, 139, 515, \dodoi{10.1093/mnras/139.4.515}

\bibitem[{{Readhead}(1980)}]{1980IAUS...92..165R}
{Readhead}, A.~C.~S. 1980, in Objects of High Redshift, ed. G.~O. {Abell} \& P.~J.~E. {Peebles}, Vol.~92, 165--175

\bibitem[{{Readhead} {et~al.}(1978){Readhead}, {Cohen}, {Pearson}, \& {Wilkinson}}]{1978Natur.276..768R}
{Readhead}, A.~C.~S., {Cohen}, M.~H., {Pearson}, T.~J., \& {Wilkinson}, P.~N. 1978, \nat, 276, 768, \dodoi{10.1038/276768a0}

\bibitem[{{Readhead} {et~al.}(1996){Readhead}, {Taylor}, {Xu}, {Pearson}, {Wilkinson}, \& {Polatidis}}]{1996ApJ...460..612R}
{Readhead}, A.~C.~S., {Taylor}, G.~B., {Xu}, W., {et~al.} 1996, \apj, 460, 612, \dodoi{10.1086/176996}

\bibitem[{{Readhead} {et~al.}(1994){Readhead}, {Xu}, {Pearson}, {Wilkinson}, \& {Polatidis}}]{1994cers.conf...17R}
{Readhead}, A.~C.~S., {Xu}, W., {Pearson}, T.~J., {Wilkinson}, P.~N., \& {Polatidis}, A.~G. 1994, in Compact Extragalactic Radio Sources, ed. J.~A. {Zensus} \& K.~I. {Kellermann}, 17

\bibitem[{{Readhead} {et~al.}(2021){Readhead}, {Kiehlmann}, {Lister}, {O'Neill}, {Pearson}, {Sheldahl}, {Siemiginowska}, {Taylor}, \& {Wilkinson}}]{2021AN....342.1185R}
{Readhead}, A. C.~S., {Kiehlmann}, S., {Lister}, M.~L., {et~al.} 2021, Astronomische Nachrichten, 342, 1185, \dodoi{10.1002/asna.20210049}

\bibitem[{{Readhead} {et~al.}(2024){Readhead}, {Ravi}, {Blandford}, {Sullivan}, {Somalwar}, {Begelman}, {Birkinshaw}, {Liodakis}, {Lister}, {Pearson}, {Taylor}, {Wilkinson}, {Globus}, {Kiehlmann}, {Lawrence}, {Murphy}, {O'Neill}, {Pavlidou}, {Sheldahl}, {Siemiginowska}, \& {Tassis}}]{2024ApJ...961..242R}
{Readhead}, A.~C.~S., {Ravi}, V., {Blandford}, R.~D., {et~al.} 2024, \apj, 961, 242, \dodoi{10.3847/1538-4357/ad0c55}

\bibitem[{{Rodriguez} {et~al.}(2006){Rodriguez}, {Taylor}, {Zavala}, {Peck}, {Pollack}, \& {Romani}}]{2006ApJ...646...49R}
{Rodriguez}, C., {Taylor}, G.~B., {Zavala}, R.~T., {et~al.} 2006, \apj, 646, 49, \dodoi{10.1086/504825}

\bibitem[{{Schmidt}(1968)}]{1968ApJ...151..393S}
{Schmidt}, M. 1968, \apj, 151, 393, \dodoi{10.1086/149446}

\bibitem[{{Siemiginowska}(2009)}]{2009AN....330..264S}
{Siemiginowska}, A. 2009, Astronomische Nachrichten, 330, 264, \dodoi{10.1002/asna.200811172}

\bibitem[{{Sillanpaa} {et~al.}(1988){Sillanpaa}, {Haarala}, {Valtonen}, {Sundelius}, \& {Byrd}}]{1988ApJ...325..628S}
{Sillanpaa}, A., {Haarala}, S., {Valtonen}, M.~J., {Sundelius}, B., \& {Byrd}, G.~G. 1988, \apj, 325, 628, \dodoi{10.1086/166033}

\bibitem[{{Sobolewska} {et~al.}(2019){Sobolewska}, {Siemiginowska}, {Guainazzi}, {Hardcastle}, {Migliori}, {Ostorero}, \& {Stawarz}}]{2019ApJ...871...71S}
{Sobolewska}, M., {Siemiginowska}, A., {Guainazzi}, M., {et~al.} 2019, \apj, 871, 71, \dodoi{10.3847/1538-4357/aaee78}

\bibitem[{{Stanghellini} {et~al.}(2009){Stanghellini}, {Dallacasa}, {Venturi}, {An}, \& {Hong}}]{2009AN....330..153S}
{Stanghellini}, C., {Dallacasa}, D., {Venturi}, T., {An}, T., \& {Hong}, X.~Y. 2009, Astronomische Nachrichten, 330, 153, \dodoi{10.1002/asna.200811144}

\bibitem[{{Stanghellini} {et~al.}(1997){Stanghellini}, {O'Dea}, {Baum}, {Dallacasa}, {Fanti}, \& {Fanti}}]{1997A&A...325..943S}
{Stanghellini}, C., {O'Dea}, C.~P., {Baum}, S.~A., {et~al.} 1997, \aap, 325, 943

\bibitem[{{Sullivan} {et~al.}(2024){Sullivan}, {Blandford}, {Begelman}, {Birkinshaw}, \& {Readhead}}]{2024MNRAS.528.6302S}
{Sullivan}, A.~G., {Blandford}, R.~D., {Begelman}, M.~C., {Birkinshaw}, M., \& {Readhead}, A. C.~S. 2024, \mnras, 528, 6302, \dodoi{10.1093/mnras/stae322}

\bibitem[{{Taylor} {et~al.}(2000){Taylor}, {Marr}, {Pearson}, \& {Readhead}}]{2000ApJ...541..112T}
{Taylor}, G.~B., {Marr}, J.~M., {Pearson}, T.~J., \& {Readhead}, A.~C.~S. 2000, \apj, 541, 112, \dodoi{10.1086/309428}

\bibitem[{{Taylor} {et~al.}(1996){Taylor}, {Readhead}, \& {Pearson}}]{1996ApJ...463...95T}
{Taylor}, G.~B., {Readhead}, A.~C.~S., \& {Pearson}, T.~J. 1996, \apj, 463, 95, \dodoi{10.1086/177225}

\bibitem[{{Taylor} {et~al.}(1994){Taylor}, {Vermeulen}, {Pearson}, {Readhead}, {Henstock}, {Browne}, \& {Wilkinson}}]{1994ApJS...95..345T}
{Taylor}, G.~B., {Vermeulen}, R.~C., {Pearson}, T.~J., {et~al.} 1994, \apjs, 95, 345, \dodoi{10.1086/192101}

\bibitem[{{Tremblay} {et~al.}(2016){Tremblay}, {Taylor}, {Ortiz}, {Tremblay}, {Helmboldt}, \& {Romani}}]{2016MNRAS.459..820T}
{Tremblay}, S.~E., {Taylor}, G.~B., {Ortiz}, A.~A., {et~al.} 2016, \mnras, 459, 820, \dodoi{10.1093/mnras/stw592}

\bibitem[{{Tschager} {et~al.}(1999){Tschager}, {Schilizzi}, {Snellen}, {de Bruyn}, {Miley}, {R{\"o}ttgering}, {van Langevelde}, {Fanti}, \& {Fanti}}]{1999NewAR..43..681T}
{Tschager}, W., {Schilizzi}, R.~T., {Snellen}, I.~A.~G., {et~al.} 1999, \nar, 43, 681, \dodoi{10.1016/S1387-6473(99)00077-9}

\bibitem[{{Valtonen} {et~al.}(2016){Valtonen}, {Zola}, {Ciprini}, {Gopakumar}, {Matsumoto}, {Sadakane}, {Kidger}, {Gazeas}, {Nilsson}, {Berdyugin}, {Piirola}, {Jermak}, {Baliyan}, {Alicavus}, {Boyd}, {Campas Torrent}, {Campos}, {Carrillo G{\'o}mez}, {Caton}, {Chavushyan}, {Dalessio}, {Debski}, {Dimitrov}, {Drozdz}, {Er}, {Erdem}, {Escartin P{\'e}rez}, {Fallah Ramazani}, {Filippenko}, {Ganesh}, {Garcia}, {G{\'o}mez Pinilla}, {Gopinathan}, {Haislip}, {Hudec}, {Hurst}, {Ivarsen}, {Jelinek}, {Joshi}, {Kagitani}, {Kaur}, {Keel}, {LaCluyze}, {Lee}, {Lindfors}, {Lozano de Haro}, {Moore}, {Mugrauer}, {Naves Nogues}, {Neely}, {Nelson}, {Ogloza}, {Okano}, {Pandey}, {Perri}, {Pihajoki}, {Poyner}, {Provencal}, {Pursimo}, {Raj}, {Reichart}, {Reinthal}, {Sadegi}, {Sakanoi}, {Salto Gonz{\'a}lez}, {Sameer}, {Schweyer}, {Siwak}, {Sold{\'a}n Alfaro}, {Sonbas}, {Steele}, {Stocke}, {Strobl}, {Takalo}, {Tomov}, {Tremosa Espasa}, {Valdes}, {Valero P{\'e}rez}, {Verrecchia}, {Webb}, {Yoneda}, {Zejmo}, {Zheng}, {Telting}, {Saario},
  {Reynolds}, {Kvammen}, {Gafton}, {Karjalainen}, {Harmanen}, \& {Blay}}]{2016ApJ...819L..37V}
{Valtonen}, M.~J., {Zola}, S., {Ciprini}, S., {et~al.} 2016, \apjl, 819, L37, \dodoi{10.3847/2041-8205/819/2/L37}

\bibitem[{{Verkhodanov} {et~al.}(2005){Verkhodanov}, {Trushkin}, {Andernach}, \& {Chernenkov}}]{2005BSAO...58..118V}
{Verkhodanov}, O.~V., {Trushkin}, S.~A., {Andernach}, H., \& {Chernenkov}, V.~N. 2005, Bulletin of the Special Astrophysics Observatory, 58, 118.
\newblock \doarXiv{0705.2959}

\bibitem[{{Wall} \& {Peacock}(1985)}]{1985MNRAS.216..173W}
{Wall}, J.~V., \& {Peacock}, J.~A. 1985, \mnras, 216, 173, \dodoi{10.1093/mnras/216.2.173}

\bibitem[{{Wilkinson} {et~al.}(1994){Wilkinson}, {Polatidis}, {Readhead}, {Xu}, \& {Pearson}}]{1994ApJ...432L..87W}
{Wilkinson}, P.~N., {Polatidis}, A.~G., {Readhead}, A.~C.~S., {Xu}, W., \& {Pearson}, T.~J. 1994, \apjl, 432, L87, \dodoi{10.1086/187518}

\bibitem[{{Wilkinson} {et~al.}(2001){Wilkinson}, {Henstock}, {Browne}, {Polatidis}, {Augusto}, {Readhead}, {Pearson}, {Xu}, {Taylor}, \& {Vermeulen}}]{2001PhRvL..86..584W}
{Wilkinson}, P.~N., {Henstock}, D.~R., {Browne}, I.~W., {et~al.} 2001, \prl, 86, 584, \dodoi{10.1103/PhysRevLett.86.584}

\bibitem[{{Xu} {et~al.}(1995){Xu}, {Readhead}, {Pearson}, {Polatidis}, \& {Wilkinson}}]{1995ApJS...99..297X}
{Xu}, W., {Readhead}, A.~C.~S., {Pearson}, T.~J., {Polatidis}, A.~G., \& {Wilkinson}, P.~N. 1995, \apjs, 99, 297, \dodoi{10.1086/192189}

\end{thebibliography}
\bibliographystyle{aasjournal}

\end{document}